\def\fracm#1#2{\hbox{\large{${\frac{{#1}}{{#2}}}$}}}

\documentstyle[12pt]{article}

\skewchar\fivmi='177
\skewchar\sixmi='177
\skewchar\sevmi='177
\skewchar\egtmi='177
\skewchar\ninmi='177
\skewchar\tenmi='177
\skewchar\elvmi='177
\skewchar\twlmi='177
\skewchar\frtnmi='177
\skewchar\svtnmi='177
\skewchar\twtymi='177
\def\@magscale#1{ scaled \magstep #1}
\skewchar\fivsy='60
\skewchar\sixsy='60
\skewchar\sevsy='60
\skewchar\egtsy='60
\skewchar\ninsy='60
\skewchar\tensy='60
\skewchar\elvsy='60
\skewchar\twlsy='60
\skewchar\frtnsy='60
\skewchar\svtnsy='60
\skewchar\twtysy='60

\catcode`@=11
\def\un#1{\relax\ifmmode\@@underline#1\else
        $\@@underline{\hbox{#1}}$\relax\fi}
\catcode`@=12

\def\a{\alpha}
\def\b{\beta}

\def\d{\delta}
\def\e{\epsilon}

\def\g{\gamma}

\def\l{\lambda}
\def\m{\mu}

\def\p{\pi}

\def\s{\sigma}
\def\t{\tau}

\def\z{\zeta}

\def\L{\Lambda}
\def\O{\Omega}

\def\dslash{\not{\hbox{\kern-2pt $\partial$}}}
\def\Dslash{\not{\hbox{\kern-4pt $D$}}}
\def\pslash{\not{\hbox{\kern-2.3pt $p$}}}
 \newtoks\slashfraction
 \slashfraction={.13}
 \def\slash#1{\setbox0\hbox{$ #1 $}
 \setbox0\hbox to \the\slashfraction\wd0{\hss \box0}/\box0 }

\font\ro=cmsy10                          
\def\kcr{{\hbox{\ro \char'170}}}                
\def\ktl{{\hbox{\ro \char'170}}}        
\def\ktr{{\hbox{\ro \char'170}}}        
\def\kbl{{\hbox{\ro \char'170}}}        
\def\kbr{{\hbox{\ro \char'170}}}        

\def\plpl{\raise-2pt\hbox{$\raise3pt\hbox{$_+$}\hskip-6.67pt\raise0.0pt
\hbox{$^+$}\hskip 0.01pt$}}
\def\mimi{\raise-2pt\hbox{$\raise3pt\hbox{$_-$}\hskip-6.67pt\raise0.0pt
\hbox{$^-$}\hskip 0.01pt$}}

\def\bo{{\raise.15ex\hbox{\large$\Box$}}}               
\def\pa{\partial}                                       
\def\su{\sum}                                           
\def\TH{{\raise.2ex\hbox{$\displaystyle \bigodot$}\mskip-4.7mu \llap H \;}}
\def\face{{\raise.2ex\hbox{$\displaystyle \bigodot$}\mskip-2.2mu \llap {$\ddot
        \smile$}}}                                      

\def\pp{{\mathchoice
          {
              \kern 1pt%
              \raise 1pt
              \vbox{\hrule width5pt height0.4pt depth0pt
                    \kern -2pt
                    \hbox{\kern 2.3pt
                          \vrule width0.4pt height6pt depth0pt
                          }
                    \kern -2pt
                    \hrule width5pt height0.4pt depth0pt}%
                    \kern 1pt
           }
            {
              \kern 1pt%
              \raise 1pt
              \vbox{\hrule width4.3pt height0.4pt depth0pt
                    \kern -1.8pt
                    \hbox{\kern 1.95pt
                          \vrule width0.4pt height5.4pt depth0pt
                          }
                    \kern -1.8pt
                    \hrule width4.3pt height0.4pt depth0pt}%
                    \kern 1pt
            }
            {
              \kern 0.5pt%
              \raise 1pt
              \vbox{\hrule width4.0pt height0.3pt depth0pt
                    \kern -1.9pt  
                    \hbox{\kern 1.85pt
                          \vrule width0.3pt height5.7pt depth0pt
                          }
                    \kern -1.9pt
                    \hrule width4.0pt height0.3pt depth0pt}%
                    \kern 0.5pt
            }
            {
              \kern 0.5pt%
              \raise 1pt
              \vbox{\hrule width3.6pt height0.3pt depth0pt
                    \kern -1.5pt
                    \hbox{\kern 1.65pt
                          \vrule width0.3pt height4.5pt depth0pt
                          }
                    \kern -1.5pt
                    \hrule width3.6pt height0.3pt depth0pt}%
                    \kern 0.5pt
            }
        }}

\def\sp#1{{}^{#1}}                              
\def\Tilde#1{\widetilde{#1}}                    
\def\Hat#1{\widehat{#1}}                        
\def\leftrightarrowfill{$\mathsurround=0pt \mathord\leftarrow \mkern-6mu
        \cleaders\hbox{$\mkern-2mu \mathord- \mkern-2mu$}\hfill
        \mkern-6mu \mathord\rightarrow$}
\def\dvec#1{\vbox{\ialign{##\crcr
        \leftrightarrowfill\crcr\noalign{\kern-1pt\nointerlineskip}
        $\hfil\displaystyle{#1}\hfil$\crcr}}}           

\def\fracm#1#2{\hbox{\large{${\frac{{#1}}{{#2}}}$}}}
\def\frac#1#2{{\textstyle{#1\over\vphantom2\smash{\raise.20ex
        \hbox{$\scriptstyle{#2}$}}}}}                   
\def\sfrac#1#2{{\vphantom1\smash{\lower.5ex\hbox{\small$#1$}}\over
        \vphantom1\smash{\raise.4ex\hbox{\small$#2$}}}} 
\def\bfrac#1#2{{\vphantom1\smash{\lower.5ex\hbox{$#1$}}\over
        \vphantom1\smash{\raise.3ex\hbox{$#2$}}}}       
\def\afrac#1#2{{\vphantom1\smash{\lower.5ex\hbox{$#1$}}\over#2}}    
\def\on#1#2{\mathop{\null#2}\limits^{#1}}               
\def\bvec#1{\on\leftarrow{#1}}                  

\newskip\humongous \humongous=0pt plus 1000pt minus 1000pt
\def\caja{\mathsurround=0pt}
\def\eqalign#1{\,\vcenter{\openup2\jot \caja
        \ialign{\strut \hfil$\displaystyle{##}$&$
        \displaystyle{{}##}$\hfil\crcr#1\crcr}}\,}
\newif\ifdtup

\def\ref#1{$\sp{#1)}$}

\parskip=\medskipamount                 
\thicklines                         

\def\oldheadpic{                                
        \setlength{\unitlength}{.4mm}
        \thinlines
        \par
        \begin{picture}(349,16)
        \put(325,16){\line(1,0){4}}
        \put(330,16){\line(1,0){4}}
        \put(340,16){\line(1,0){4}}
        \put(335,0){\line(1,0){4}}
        \put(340,0){\line(1,0){4}}
        \put(345,0){\line(1,0){4}}
        \put(329,0){\line(0,1){16}}
        \put(330,0){\line(0,1){16}}
        \put(339,0){\line(0,1){16}}
        \put(340,0){\line(0,1){16}}
        \put(344,0){\line(0,1){16}}
        \put(345,0){\line(0,1){16}}
        \put(329,16){\oval(8,32)[bl]}
        \put(330,16){\oval(8,32)[br]}
        \put(339,0){\oval(8,32)[tl]}
        \put(345,0){\oval(8,32)[tr]}
        \end{picture}
        \par
        \thicklines
        \vskip.2in}
\def\oldtitle#1#2#3#4{\oldheadpic\begin{center}\vglue.5in{\large\bf #1}\\[.6in]
        {#2}\\[.1in] {\it Department of Physics and Astronomy}\\
        {\it University of Maryland, College Park, MD 20742}\\[.6in]
        Physics Publication \#{#3}\\ {#4}\\[1.5in] {\bf ABSTRACT}\\[.1in]
        \end{center} \begin{quotation}}                 
\def\oldTitle#1#2#3#4#5#6#7{\oldheadpic\begin{center} \vglue .4in
        {\large\bf #1}\\[.4in]
        {#2}\\[.1in] {\it Department of Physics and Astronomy}\\
        {\it University of Maryland, College Park, MD 20742}\\[.1in]
        {#3}\\[.1in] {\it {#4}}\\ {\it {#5}}\\[.4in]
        Physics Publication \#{#6}\\ {#7}\\[.5in] {\bf ABSTRACT}\\[.1in]
        \end{center} \begin{quotation}}                 
\def\border{                                            
        \setlength{\unitlength}{1mm}
        \newcount\xco
        \newcount\yco
        \xco=-21
        \yco=12
        \begin{picture}(140,0)
        \put(\xco,\yco){$\ktl$}
        \advance\yco by-1
        {\loop
        \put(\xco,\yco){$\kcr$}
        \advance\yco by-2
        \ifnum\yco>-240
        \repeat
        \put(\xco,\yco){$\kbl$}}
        \xco=158
        \yco=12
        \put(\xco,\yco){$\ktr$}
        \advance\yco by-1
        {\loop
        \put(\xco,\yco){$\kcr$}
        \advance\yco by-2
        \ifnum\yco>-240
        \repeat
        \put(\xco,\yco){$\kbr$}}
        \put(-20,13){\tiny University of Maryland Elementary Particle
Physics University of Maryland Elementary Particle Physics University of
Maryland Elementary Particle Physics}
        \put(-20,-241.5){\tiny University of Maryland Elementary
Particle Physics University of Maryland Elementary Particle Physics
University of Maryland Elementary Particle Physics}
        \end{picture}
        \par\vskip-8mm}
\def\bordero{                                           
        \setlength{\unitlength}{1mm}
        \newcount\xco
        \newcount\yco
        \xco=-31
        \yco=12
        \begin{picture}(140,0)
        \put(\xco,\yco){$\ktl$}
        \advance\yco by-1
        {\loop
        \put(\xco,\yco){$\kclr}
        \advance\yco by-2
        \ifnum\yco>-240
        \repeat
        \put(\xco,\yco){$\kbl$}}
        \xco=151
        \yco=12
        \put(\xco,\yco){$\ktr$}
        \advance\yco by-1
        {\loop
        \put(\xco,\yco){$\kcr$}
        \advance\yco by-2
        \ifnum\yco>-240
        \repeat
        \put(\xco,\yco){$\kbr$}}
        \put(-20,12){\ooo
bacdefghidfghghdhededbihdgdfdfhhdheidhdhebaaahjhhdahba

hgdedge
   hgfdiehhgdigicba}
        \put(-20,-241.5){\ooo
ababaighefdbfghgeahgdfgafagihdidihiidhiagfedhadbfd

ecdcdfa
   gdcbhaddhbgfchbgfdacfediacbabab}
        \end{picture}
        \par\vskip-8mm}
\def\headpic{                                           
        \indent
        \setlength{\unitlength}{.4mm}
        \thinlines
        \par
        \begin{picture}(29,16)
        \put(165,16){\line(1,0){4}}
        \put(170,16){\line(1,0){4}}
        \put(180,16){\line(1,0){4}}
        \put(175,0){\line(1,0){4}}
        \put(180,0){\line(1,0){4}}
        \put(185,0){\line(1,0){4}}
        \put(169,0){\line(0,1){16}}
        \put(170,0){\line(0,1){16}}
        \put(179,0){\line(0,1){16}}
        \put(180,0){\line(0,1){16}}
        \put(184,0){\line(0,1){16}}
        \put(185,0){\line(0,1){16}}
        \put(169,16){\oval(8,32)[bl]}
        \put(170,16){\oval(8,32)[br]}
        \put(179,0){\oval(8,32)[tl]}
        \put(185,0){\oval(8,32)[tr]}
        \end{picture}
        \par\vskip-6.5mm
        \thicklines}
\def\title#1#2#3#4{\border\headpic {\hbox to\hsize{#4 \hfill UMDEPP #3}}\par
        \begin{center} \vglue .5in {\large\bf #1}\\[.6in]
        {#2}\\[.1in] {\it Department of Physics and Astronomy}\\
        {\it University of Maryland, College Park, MD 20742}\\[1.5in]
        {\bf ABSTRACT}\\[.1in] \end{center} \begin{quotation}}  
\def\Title#1#2#3#4#5#6#7{\border\headpic
        {\hbox to\hsize{#7 \hfill UMDEPP #6}}\par
        \begin{center} \vglue .4in {\large\bf #1}\\[.4in]
        {#2}\\[.1in] {\it Department of Physics and Astronomy}\\
        {\it University of Maryland, College Park, MD 20742}\\[.1in]
        {#3}\\[.1in] {\it {#4}}\\ {\it {#5}}\\[.5in] {\bf ABSTRACT}\\[.1in]
        \end{center} \begin{quotation}}                 
\def\endtitle{\end{quotation}\newpage}                  

\def\qd{{\kern0.5pt
                   q \kern-5.05pt \raise5.8pt\hbox{$\textstyle.$}\kern
0.5pt}}

\begin{document}

\def\gfrac#1#2{\frac {\scriptstyle{#1}}
        {\mbox{\raisebox{-.6ex}{$\scriptstyle{#2}$}}}}
\def\gg{{\hbox{\sc g}}}
\border\headpic {\hbox to\hsize{October 1995 \hfill {UMDEPP 96-39}}}
\par
\setlength{\oddsidemargin}{0.3in}
\setlength{\evensidemargin}{-0.3in}
\begin{center}
\vglue .08in
{\large\bf A Proposal for ${\aleph}_0$ \\
Extended Supersymmetry in\\
Integrable Systems\footnote {Supported in part by National
Science Foundation Grant PHY-94-21386 \newline ${~~~~~}$ and by NATO
Grant CRG-93-0789}  }
\\[.72in]

S. James Gates, Jr. and Lubna Rana
\\[.02in]
{\it Department of Physics\\
University of Maryland\\
College Park, MD 20742-4111  USA}\\[.2in]
{\bf {\tt gates@umdhep.umd.edu}}\\
{\bf {\tt lubna@umdhep.umd.edu}}\\[3in]

{\bf ABSTRACT}\\[.002in]
\end{center}
\begin{quotation}
{Utilizing techniques suggested by the recently obtained construction
of off-shell spinning particles, we propose the arbitrary $N$-extension
of supersymmetry for the KdV system. It is further suggested that the
${\aleph}_0$ extension for the SKdV system provides a paradigm
for {\underline {all}} supersymmetric completely integrable systems.}

\endtitle
\section{Introduction}

{}~~~~The topic of integrable or completely solvable systems is one with
a long history having perhaps its best known origin in an observation
of John Scott Russell\footnote{In a somewhat jocular way, we may say
this was, perhaps the first experimental observation of \newline
${~~~~~}$ supersymmetry in Nature.} in 1834 \cite{A}.  Almost sixty
years later the mathematical setting for this class of theories
was established \cite{B}.  Finally one-hundred and fifty years later,
explorations of the connection between supersymmetry and integrable
systems began in earnest \cite{C,D}. The topic of integrable systems
has also been found to coalesce with relativistic particle and spinning
particles \cite{E} in an unexpected way wherein the KdV and SKdV Lax
operators can be found by use of an appropriate set of variables \cite{F}.
Due to this last observation, advances in our understanding of the
spinning particle quite naturally should have consequences for our
understanding of SKdV systems. Along this line of thought, we
\cite{G} have recently been able to give, for the first time, an
off-shell description of the spinning particle for arbitrary $N$,
the degree of the supersymmetry extension. Since $N$ is an arbitrarily
large integer, the set of all such integers constitutes a representation
of $\aleph_0$, the ``smallest'' transfinite number.

In the present brief note, we wish to show that the off-shell momentum
multiplet of the $\aleph_0$ supersymmetric spinning particle appears
to provide the fundamental supersymmetric representation for the
construction of the supersymmetric extension of the KdV equation.
We show that the well known cases of the $N$ = 1 and $N$ = 2 \cite{D}
theories are ``naturally'' embedded in a simple algebraic structure.
Extending this embedding to the entire structure suggests a form
for the supersymmetric extension for arbitrary values of $N$.  We
discuss the cases of $N$ = 3, 4 and compare to the suggestion of Delduc
and Ivanov \cite{G1} for the SKdV system. We end our letter with a
conjecture that the structure we have found is universal for all
supersymmetric integrable systems.

\section{A Universal Supersymmetry Representation for Integrable
Systems}

{}~~~~In a related work \cite{H}, we have shown that associated with
the off-shell spinning particle coordinate there is a ``momentum
multiplet.'' One such multiplet occurs for each coordinate of the
spinning particle. After a certain transformation, the component
fields $(\, w_i {}^j (x,t), \,  \xi_{\rm I} (x,t), \, \xi_i {}^{\hat
k} (x,t), \, u (x,t) \,)$ of the multiplet have supersymmetry
variations given
by
$$ \eqalign{ {~~~~~~~~~}
{\d}_{Q} \,  w_{i} {}^{j} &=~ - i \, 2 \, \a_{\rm I} \, [~ \left(
f_{{\rm I} \, {\rm J}} \right)_{i} {}^{j} \,  \xi_{\rm J}  \, + \,
\left( {\rm R}_{\rm I} \right)_{\hat k} {}^{j} \, \xi_i {}^{\hat k}
{}~]   ~~~~, \cr
{\d}_{Q} \, \xi_{\rm I}  &=~   \a_{\rm I} \, u ~+~ \, {\rm d}^{-1}
\a_{\rm J} \left( f_{{\rm I}\, {\rm J}} \right)_j {}^{i} \, \pa_x
w_{i} {}^{j} ~~~~, \cr
{\d}_{Q} \, \xi_i {}^{\hat k} &=~ -  \, \a_{\rm I} \, [~ \left({\rm
L}_{\rm I}  \right)_{j} {}^{\hat k} \, \pa_x w_{i} {}^{j} \, + \,
{\rm d}^{-1} \left( {\rm L}_{\rm J} \right)_{i} {}^{\hat k} \, \left(
f_{{\rm I} \, {\rm J}} \right)_j {}^{l} \, \pa_x  w_{l} {}^j  ~]
{}~~~~, \cr
{\d}_{Q} \, u &=~ -i \, 2 \a_{\rm I} \, \pa_x \xi_{\rm I} ~~~~.
} \eqno(2.1) $$
here ${\rm d} = 2$, $w_i {}^i = \left( {\rm R}_{\rm I} \right)_{\hat k} {}^i
\xi_i {}^{\hat k} = 0$. (See \cite{G} for notational conventions as well
as the appendix.)   For our purposes, it is also useful to introduce the
following decomposition $ w_{i} {}^{j} \equiv \left( f_{{\rm I} \, {\rm J}}
\right)_{i} {}^{j} w_{{\rm I} \, {\rm J}} ~+~ {\Hat w}_{i} {}^{j}$ and
$ \left( f_{{\rm I} \, {\rm J}} \right)_{j} {}^{i} {\Hat w}_{i} {}^{j}
\equiv 0$. The field $u$ corresponds to a momentum of a single coordinate
of the spinning particle.

Alternately, we may take the 3D, $\aleph_0$ supersymmetric abelian Yang-Mills
multiplet with field content (${\cal B}_i {}^j$, $\l_{\a \, \hat k}
{}^i$, $A_a$) and supersymmetry variations \cite{G},
$$\eqalign{
\d_Q {\cal B}_i {}^j &=~  \e^{\a \, {\rm I}} \, ({\rm L}_{\rm I})_k
{}^{\hat k} \, \left[ ~ \d_i {}^k \l_{\a \, \hat k} {}^j ~-~ {\rm
d}^{-1} \d_i {}^j \l_{\a \, \hat k} {}^k   ~ \right] ~~~, \cr
\d_Q \l_{\a \, \hat k} {}^k &=~ i \e^{\b \, {\rm I}} \, ({\rm
R}_{\rm I} )_{\hat k} {}^j (\g^a )_{\a \b} ~\left[ \, \pa_a {
\cal B}_j {}^k ~+~ \fracm 12 \, {\rm d}^{-1} \, \d_j {}^k \e_a
{}^{b c} F_{b c}  \, \right] ~~~, \cr
\d_Q A_a &=~ i \e^{\a \, {\rm I}} \, ({\rm L}_{\rm I})_k {}^{\hat
k}  \, (\g_a )_{\a \b}  \, \l^{\b}{}_{  \hat k} {}^k   ~~~. }
\eqno(2.2)$$
(where ${\cal B}_i {}^i  = 0$), as a starting point. We next separate
the gaugini according to the definition
$$\l_{\a \, \hat k} {}^i ~=~ {\rm d}^{-1} \, \Big[ ~ ({\rm R}_{\rm
I} )_{\hat k} {}^i \l_{\rm I} ~+~ {\Hat \l}_{\a \, \hat k} {}^i ~
\Big] ~~~,~~~ ({\rm L}_{\rm I})_i {}^{\hat k} {\Hat \l}_{\a \, \hat
k} {}^i ~=~ 0 ~~~~,  \eqno(2.3)$$
where upon the variations in (2.2) take the forms,
$$ \eqalign{ {~~~~~~}
\d_Q {\cal B}_i {}^j &=~ \e^{\a \, {\rm I}}  \left[ ~  \left( f_{{
\rm I} \, {\rm J}} \right)_{i} {}^{j} \,  \l_{\a \, {\rm J}}  \, + \,
\left( {\rm L}_{\rm I}  \right)_{i} {}^{\hat k} \, {\Hat \l}_{\a \,
\hat k} {}^j ~ \right] ~~~, \cr
\d_Q \l_{\a \, {\rm I}} &=~ i \, \e^{\b \, {\rm J}} (\g^a)_{\a \b}
\left[ ~ \frac 12 \d_{\rm {I \, J}} \e_a {}^{ b \, c} F_{ b \, c} ~-~
{\rm d}^{-1} \, \left( f_{{\rm I} \, {\rm J}} \right)_{i} {}^{j} ( \,
\pa_a {\cal B}_j {}^i \,) ~ \right] ~~~, \cr
\d_Q {\Hat \l}_{\a \, \hat k} {}^k &=~  i \, \e^{\b \, {\rm J}} (
\g^a)_{\a \b}  \left[ ~ ({\rm R}_{\rm I} )_{\hat k} {}^i ( \, \pa_a
{\cal B}_i {}^k \,)  ~-~ {\rm d}^{-1} \, ({\rm R}_{\rm I} )_{\hat k}
{}^k \left( f_{{\rm I} \, {\rm J}} \right)_{i} {}^{j} ( \, \pa_a {\cal
B}_j {}^i \,) ~ \right] ~~~, \cr
\d_Q A_a &=~ - \, i \e^{\a \, {\rm I}} \,  (\g_a )_{\a \b}
\, \l^{\b}{}_{\rm I} ~\to ~ \d_Q \Big( \, \e_{a \, b \, c} F^{ b \, c}
\, \Big) ~=~  - i 2 \e^{\a \, {\rm I}} \e_{a \, b \, c} (\g^b)_{\a \b}
\pa^c  \l^{\b}{}_{\rm I} ~~~, }
\eqno(2.4)$$
after rescaling ${\cal B}_i {}^j \to {\rm d}^{-1} {\cal B}_i {}^j$. Next
we perform a reduction from 3D to 1D defined by
$$
\pa_a ~\to~ ( 0, \, \pa_x , \, 0) ~~~,~~~ A_a (t, \, x , \, y) ~\to~
( 0 , \, 0 , \, A_y(x) ) ~~~,  ~~~ \e_{a \, b \, c} F^{b \, c} \to
2\, ( \pa_x A_y , \, 0 , \, 0) ~~~,
\eqno(2.5)$$
and demand the consistency of the condition $\d_Q A_t = \d_Q A_x = 0$.
These consistency conditions lead to 1D spinors (i.e. solutions take
the forms $\l_{\a}{}^{\rm I} = a u_{\a}(-) \xi^{\rm I}, ~ \l_{\a \,
\hat k}{}^k =  b u_{\a}(+) \xi_{\hat k}{}^k$ where $u_{\a}(\pm)$ are
the eigenspinors for $(\g_{y})_{\a}{}^{\b}$ and $a$ and $b$ are constants)
that when substituted back into (2.4) yield a set of transformations
equivalent to (2.1) with the mappings $\pa_x A_y \to u , \, {\cal B}_i
{}^j \to w_i {}^j, \,  \l_{\a \, {\rm I}} \to \xi_{\rm I}$ and ${\Hat
\l}_{\a \, \hat k} {}^k \to \xi_i {}^{\hat k}$.  So both the $\aleph_0$
spinning particle as well as the 3D $\aleph_0$ supersymmetric abelian
Yang-Mills multiplet lead to the same result.

In order to see how (2.1) is related to the standard known constructions
of $N$ = 1 and $N$ = 2 SKdV \cite{D}, we define $\left( {\rm L}_{\rm I}
\right)_{i} {}^{\hat k} = (\,- {\rm I}, \, i \s^{\Hat  2} \,)$, $\left(
{\rm R}_{\rm I} \right)_{\hat k} {}^i = (\, {\rm I}, \, i \s^{\Hat  2}
\,)$ and $\left( f_{{\rm I} \, {\rm J}} \right)_{i}{}^{j} = - i \, \e_{
{\rm I} \, {\rm J}} \left( \s^{\Hat  2} \right)_{i} {}^{j}$ where $\left(
\s^{\Hat 2} \right)_{i} {}^{j}$ denotes the usual second $2 \times 2$
Pauli matrix. From the constraints on $w_i {}^j$ and $\xi_i {}^{\hat k}$
it follows that
$$  \eqalign{ {~~~~~~~~~}
w_i {}^j &=~ w \left( i \s^{\Hat 2} \right)_{i} {}^{j} ~+~
w^{\Hat 3} \left( \s^{\Hat 3} \right)_{i} {}^{j} ~+~ w^{\Hat 1}
\left( \s^{\Hat 1} \right)_{i} {}^{j} ~~~~, \cr
&\equiv ~  w \left( i \s^{\Hat 2} \right)_{i} {}^{j} ~+~
{\Hat w}_i {}^j  ~~~~,  \cr
\xi_i {}^{\hat k} &=~ \xi^{\Hat 1} \left( \s^{\Hat 1} \right)_{i}
{}^{\hat k} ~+~ \xi^{\Hat 3} \left( \s^{\Hat 3} \right)_{i}
{}^{\hat k} ~~~~. }
\eqno(2.6) $$
In general the off-shell representation in (2.1) is reducible.  This
feature is seen by writing out the supersymmetry variations of (2.1)
in terms of the variables defined in (2.6).   A primary irreducible
submultiplet is provided by $(w , \, \xi_{\rm I} ,\, u)$ and a secondary
irreducible submultiplet is provided by
$( w^{\Hat 1} , \, w^{\Hat 3} , \, \xi^{\Hat 1} , \, \xi^{\Hat 3} )$.
They have the respective transformations laws,
$$ \eqalign{ {~~~~~~~~~}
{\d}_{Q} \, { w}  &=~  i\,  2  \a_{\rm I} \e_{{\rm I} \, {\rm J}} \xi_{\rm J}
{}~~~~, \cr
{\d}_{Q} \, \xi_{\rm I}  &=~   \a_{\rm I} \, u ~+~   \, \a_{\rm J} \, \e_{
\rm {I J}} \pa_x w ~~~~, \cr
{\d}_{Q} \, u &=~ -i \, 2 \a_{\rm I} \, \pa_x \xi_{\rm I} ~~~~,}
\eqno(2.7) $$
and separately the transformations laws,
$$ \eqalign{ {~~~~~~~~~}
{\d}_{Q} \, {\Hat w}_{i} {}^{j} &=~ - i \, 2  \, \a_{\rm I} \left( {\rm
R}_{\rm I} \right)_{\hat k} {}^{j} \, \xi_i {}^{\hat k}   ~~~~, \cr
{\d}_{Q} \, \xi_i {}^{\hat k} &=~ -  \, \a_{\rm I} \, [~  \left({\rm L
}_{\rm I} \right)_{j} {}^{\hat k} \, \pa_x {\Hat w}_{i} {}^{j} \, + \,
{\rm d}^{-1} \left( {\rm L}_{\rm J} \right)_i {}^{\hat k} \, \left(
f_{{\rm I} \, {\rm J}} \right)_j {}^{l} \, \pa_x {\Hat w}_{l} {}^j ~]  ~~~~.
}\eqno(2.8) $$

The reader familiar with the literature of the SKdV equation will immediately
recognize (2.7) as a form of the supersymmetric multiplet that is known
to occur in SKdV theories. In order to impose the condition that the
multiplet of (2.1) should obey the dynamics of the SKdV equations, we must
impose two conditions
$$  \eqalign{ {~~~~~~~~~}
0 &=~ \left( f_{{\rm I} \, {\rm J}} \right)_{j} {}^{i} \Big[ ~  \pa_t \,
{w}_{i} {}^{j} ~+~  \pa_x^3 {w}_{i} {}^{j} ~+~ 6 {\rm d}^{-1}  {w}_k {}^l
{w}_l {}^k \pa_x {w}_{i} {}^{j} ~-~ 3 \, \pa_x ( u {w}_{i} {}^{j} ) ~
\Big] ~~~~, \cr
0 &=~ {\Hat w}_{j} {}^{k} ~~~~.  }\eqno(2.9) $$
The first supersymmetry variation of these yield,
$$  \eqalign{ {~~~~~~~~~}
0 &=~ \pa_t \xi_{\rm I} ~+~ \pa_x^3 \xi_{\rm I} ~-~ 6 \pa_x (\,
w^2 \xi_{\rm I} \,) ~-~ 3 \pa_x (\, u \xi_{\rm I} \,)
{}~-~ 3 \e_{{\rm I} \, {\rm J}} \pa_x (\, w  \pa_x \xi_{\rm J} \,)
{}~~~~, \cr
0 &=~ \xi_i {}^{\hat k} ~~~~.  }\eqno(2.10) $$
Next and finally the second supersymmetry variation yields
$$  \eqalign{ {~~~~~~~~~}
0 &=~  \pa_t u ~+~ \pa_x^3 u ~-~ 6 u \pa_x u  ~-~ 6 \pa_x (\,  u w^2  \,)
{}~+~ 3 \pa_x (\, w \pa_x^2 w \,) {~~~~~~~~~} \cr
&{~~~}~-~ i \, 6 \pa_x (\, \xi_{\rm I}
\pa_x \xi_{\rm I} \,)  ~+~  i \, 12 \,  \e_{{\rm I} \, {\rm J}}  \pa_x
(\, w \xi_{\rm I}  \xi_{\rm J} \,)  ~~~~, \cr
0 &=~ \pa_x {\Hat w}	_i {}^{k} ~~~~.  }\eqno(2.11) $$
The second equation of (2.11) is already satisfied due to the second
equation of (2.9). The first equation of (2.11) is just the
supersymmetrically extended version of the KdV equation\footnote{
It is interesting to note that the first equation of (2.9) corresponds
to the choice $a = 1$ of \newline ${~~~~~}$ reference \cite{D}.  This
is the only value of this parameter that is consistent with the matrix
structure \newline ${~~~~~}$ of (2.1)}.

The equations (2.9), (2.10) and (2.11) can be summarized very succinctly in
terms of superfields. Let ${\cal A}_{{\rm I} \, {\rm J}}$ be defined by the
rhs of the first line of (2.9). Similarly introduce a superfield $\O_i{}^j$
whose lowest component corresponds to the second line of (2.9). Introduce a
superspace covariant derivative denoted by ${D}_{\rm I}$. The first lines of
(2.9), (2.10) and (2.11) then take the respective forms ${\cal A}_{{\rm I}
\, {\rm J}} = 0$, ${D }_{\rm J} {\cal A}_{{\rm I} \, {\rm J}} = 0$ and ${D
}_{\rm I} {D}_{\rm J} {\cal A}_{{\rm I} \, {\rm J}} = 0$.  The second lines
of (2.9), (2.10) and (2.11) then take the respective forms $\O_i {}^j = 0$,
${D }_{\rm J} \O_i {}^j = 0$ and ${D}_{\rm I} {D}_{\rm J} \O_i {}^j = 0$.

\section{ ${\aleph}_0$ Supersymmetry and the Korteweg de Vries \newline
${\,}$ Equation}

{}~~~~It may seem that so far, all we have done is a recapitulation of
the standard and well established SKdV system. In fact, we have much
much more. This is implicit in the seemingly strange notation in which
we cast our beginning. The point is that the form of the $2 \times 2$
Pauli matrices is dictated if we view these as a representation of a
general algebraic structure that we denote by ${\cal G}{\cal R}$(${\rm
d}, N$) (dimension ${\rm d}$ and rank $N$).   Any matrix
representation consists of $N$ linearly independent, ${\rm d} \times
{\rm d}$, real matrices (denoted by ${\rm L}_{\rm I}$)
that satisfy a general real ($\equiv {\cal G}{\cal R}$) Pauli
algebra
$$ {\rm L}_{{\rm I}} \, {\rm R}_{{\rm J}}  ~ + ~  {\rm L}_{\rm J} \,
{\rm R}_{{\rm I}} ~~=~~- \, 2 \,  \d_{{\rm I}\, {\rm J}} \, {\rm I}
{}~~~~, ~~~~ {\rm R}_{{\rm I}} \, {\rm L}_{{\rm J}}  ~ + ~  {\rm R}_{{\rm J}}
{\rm L}_{{\rm I}}  ~~=~~-  \, 2 \, \d_{{\rm I}\, {\rm J}} \, {\rm I}
{}~~~~,
\eqno(3.1) $$
where the $({\rm R}_{\rm I})$ matrices are defined by $  ({\rm L}_{\rm I}
)_{i \hat k} + ({\rm R}_{\rm I} )_{\hat k i} = 0$, ${\rm I} = 1,..., N$ and
$i, \, \hat k = 1,...,{\rm d}$.  We emphasize  that our ${\rm L}$ and
${\rm R}$ matrices are to be manipulated using Van der Waerden techniques.
For our later convenience, we define the ``complex structure'' matrices
associated with the ${\cal G} {\cal R}$(${\rm d}, N$) algebras by
$(f_{{\rm I}\, {\rm J}})_i {}^j \equiv \fracm 12 ( {\rm L}_{\rm I} {\rm
R}_{\rm J} - {\rm L}_{\rm J} {\rm R}_{\rm I})_i {}^j$ and $({\Tilde f}_{{
\rm I}\, {\rm J}})_{\hat k} {}^{\hat l} \equiv  \fracm 12 ( {\rm R}_{\rm I}
{\rm L}_{\rm J} - {\rm R}_{\rm J} {\rm L}_{\rm I})_{\hat k} {}^{\hat l}$.

An explicit representation of these ${\cal G}{\cal R}$(${\rm d}, N$)
algebras can be found in our works \cite{G,H}. The first discussion of
this type of real Clifford algebra was given by Okubo \cite{I} and
this algebraic structure has also been noted in a study of 3D non-linear
$\s$-models \cite{I1}.

One additional algebraic structure that we find useful to introduce is
what we call ${\cal U}{\cal G}{\cal R}$ defined by
$${\cal U}{\cal G}{\cal R} ~=~ \su_N \oplus {\cal G} {\cal R}(N) ~~~~.
\eqno(3.2) $$

It is a simple matter to show that the transformations in (2.1) close
uniformly on all of the fields
$$
\left[ ~ \d_{Q} \left( \a_{1} \right)  \, , \, \d_{Q}\left( \a_{2}
\right) ~ \right] ~=~  i \, 4 \, \a_1 {}^{{\rm I}} \, \a_2 {}^{{\rm I}}
\pa_{\t} ~~~,
\eqno(3.3)$$
{\underline {without}} the use of equations of motion. This is a
consequence of (3.1). However, the more useful observation is that
the transformations in (2.1) are ``${\cal U}{\cal G}{\cal R}$-covariant.''
By this we mean that for each value of $N$ there exist ${\rm d} \times
{\rm d}$ matrices contained in ${\cal U}{\cal G}{\cal R}$ for which
the algebra in (3.3) can be shown to close.  Furthermore, the equations
of (2.9) are also ${\cal U}{\cal G}{\cal R}$-covariant. Thus, we may
say that (2.1) together with (2.9) defines an $\aleph_0$ supersymmetric
extension of the known SKdV equations!

\section{Proposed Extensions for N = 3, 4 SKdV}

{}~~~~In order to demonstrate the significance of the statements at the
end of the last section, we believe it is useful to explicitly show
what the ${\cal U}{\cal G}{\cal R}$-covariant formalism suggests as
$N$ = 3, 4 SKdV theories. We begin with the $N$ = 3 theory (${\rm d}
= 4$) where the ${\cal G}{\cal R}$-Van der Waerden (1,1) tensors are
denoted by $$ \left({\rm I} \right)_{i \hat k} ~~~,~~~ \left( {\rm
L}_{\rm I} \right)_{i \hat k} ~~~,~~~ \left( {\rm E}_{\rm I} \right)_{i
\hat k}  ~~~,~~~ \left( {\rm L}_{\rm I}{\rm E}_{\rm J} \right)_{i
\hat k} ~~~,
\eqno(4.1)$$
and have multiplicities 1 + 3 + 3 + 9.  The (2,0) ${\cal G}{\cal R}$-Van
der Waerden tensors have the same multiplicities and are denoted by
$$ \left({\rm I} \right)_{i j} ~~~,~~~ \left( f_{{\rm I} }
\right)_{i j} ~~~,~~~ \left( {\rm F}_{{\rm I} } \right)_{i j} ~~~,~~~
\left( f_{\rm I} {\rm F}_{\rm J} \right)_{i j} ~~~,
\eqno(4.2)$$
where $ ({\rm E}_{{\rm I}} )_{~ i \, \hat k} ({\rm E}_{ {\rm J }})_{~ j
\, \hat k} = \d_{ { {\rm I}} \, { {\rm J}} } \, \d_{ i \, j} + \e_{ {{
{\rm I}} \, { {\rm J}} \, { {\rm K}} } } ({\rm F}_{{\rm K}} )_{ i \, j} $.
The $N$ = 3 quantities $w_i {}^j$ and $\xi_i {}^{\hat k}$ take the forms
$$  \eqalign{ {~~~~~~~~~}
w_i {}^j &=~
w^{{\rm I} } \left( f_{{\rm I} } \right)_{i} {}^{j} ~+~ {\Hat w}^{{\rm
I} } \left( {\rm F}_{{\rm I} } \right)_{i} {}^{j} ~+~ {\Hat w}^{{\rm I}\,
{\rm J} } \left( f_{\rm I} {\rm F}_{\rm J} \right)_{i} {}^{j} ~~~~, \cr
&\equiv ~  w^{{\rm I} } \left( f_{{\rm I} } \right)_{i}
{}^{j} ~+~ {\Hat w}_i {}^j  ~~~~,  \cr
\xi_i {}^{\hat k} &=~ \xi \d_{i}{}^{\hat k} ~+~ {\Hat \xi}^{\rm I}
\left({\rm E}_{\rm I} \right)_{i} {}^{\hat k} ~+~ {\Hat \xi}^{{\rm I}
\, {\rm J}} \left( {\rm L}_{\rm I} {\rm E}_{\rm J} \right)_{i} {}^{\hat
k}  ~~~~. }
\eqno(4.3) $$

For the $N$ = 4 theory (${\rm d} = 4$) we must introduce two types of
indices ${\rm I}$ = 1, 2, 3, 4 and ${\Hat {\rm I}}$ = 1, 2, and 3.
The ${\cal G}{\cal R}$-Van der Waerden (1,1) tensors are denoted by
$$  \left( {\rm L}_{\rm I} \right)_{i \hat k} ~~~,~~~ \left( \,
{\rm E}_{\Hat {\rm I}} \, \right)_{i \hat k} ~~~,~~~
\left( f_{{\rm I} \, {\rm J}} {\rm E}_{\Hat {\rm K}} \, \right)_{i
\hat k} ~~~, \eqno(4.4)$$
and have multiplicities 4 + 3 + 9.  The  ${\cal G}{\cal R}$-Van der
Waerden (2,0) tensors have the multiplicities (the (0,2) tensors have
the same decomposition)
1 + 3 + 3 + 9 and are denoted by
$$ \left({\rm I} \right)_{i j} ~~~,~~~ \left( f_{{\rm I} \, {\rm J}}
\right)_{i j} ~~~,~~~ \left( {\rm F}_{\Hat {\rm K}} \right)_{i j} ~~~,~~~
\left(  f_{{\rm I} \, {\rm J}} {\rm F}_{\Hat {\rm K}} \right)_{i j} ~~~.
\eqno(4.5)$$
The tensorial quantities in (4.5) satisfy
antisymmetry and self-duality conditions
$$\left( f_{{\rm I} \, {\rm J}} \right) = - \left( f_{{\rm J} \, {\rm
I} } \right) ~~,~~  \left(  f_{{\rm I} \, {\rm J}} {\rm F}_{\Hat {\rm K}}
\right) = - \left(  f_{{\rm J} \, {\rm I}} {\rm F}_{\Hat {\rm K}} \right) ~~,
$$
$$\left( f_{{\rm I} \, {\rm J}} \right) = \frac 12 \e_{{\rm I} \, {\rm J}
{\rm K} \, {\rm L}} \left( f_{{\rm K} \, {\rm L} } \right) ~~,~~
 \left( f_{{\rm I} \, {\rm J}} {\rm F}_{\Hat {\rm K}} \right) =
\frac 12 \e_{{\rm I} \, {\rm J} {\rm K} \, {\rm L}} \left( f_{{\rm K}
\, {\rm L}}  {\rm F}_{\Hat {\rm K}} \right) ~~~.
\eqno(4.6)$$
The $N$ = 4 quantities $w_i {}^j$ and $\xi_i {}^{\hat k}$ take the forms
$$  \eqalign{ {~~~~~~~~~}
w_i {}^j &=~
w^{{\rm I} \, {\rm J}} \left( f_{{\rm I} \, {\rm J}} \right)_{i} {}^{j} ~+~
{\Hat w}^{\Hat {\rm K}} \left( {\rm F}_{\Hat {\rm K}} \right)_{i} {}^{j}
{}~+~ {\Hat w}^{{\rm I} \, {\rm J} \, {\Hat {\rm K}}} \left( f_{{\rm I} \,
{\rm J}} {\rm F}_{\Hat {\rm K}} \right)_{i} {}^{j}
  ~~~~, \cr
&\equiv ~  w^{{\rm I} \, {\rm J}} \left( f_{{\rm I} \, {\rm J}} \right)_{i}
{}^{j} ~+~ {\Hat w}_i {}^j  ~~~~,  \cr
\xi_i {}^{\hat k} &=~  {\Hat \xi}^{~ \Hat {\rm I}} \left({\rm E}_{\Hat {\rm I}}
\right)_{i}{}^{\hat k} ~+~ {\Hat \xi}^{{\rm I} \, {\rm J} \, {\Hat {\rm K}}}
\left( f_{{\rm I}\, {\rm J}} {\rm E}_{\Hat {\rm K}} \right)_{i} {}^{\hat k}
{}~~~~. }
\eqno(4.7) $$

At this stage, there are some fundamental differences between (4.3) and (4.7)
that are very important. In the $N$ = 3 case, we note that the number of
components of ${\Hat w}_i {}^j$ is equal to the sum of the numbers of ${\Hat
\xi}^{\rm I}$ and ${\Hat \xi}^{{\rm I} \, {\rm J}}$.  This is a signal that
an off-shell $N$ = 3 theory occurs if we only retain $w^{\rm I}$ and $\xi$.
The primary $N$ = 3 irreducible off-shell submultiplet consists of $( w_{\rm
I}, \, \xi_{\rm I} , \,  \xi , \, u)$. This is very different from the $N$ =
4 case.  There we see that the number of components of ${\Hat w}_i {}^j$ is
equal to the number of components of $\xi_i {}^{\hat k}$. In the $N$ = 4 case
an off-shell formulation occurs if we retain only $w^{{\rm I} \, {\rm J}}$.
The primary $N$ = 4 irreducible off-shell multiplet consists of $(w_{{\rm I}
\, {\rm J}} ,  \,  \xi_{\rm I} , \, u )$. The primary $N$ = 3 and $N$ = 4
submultiplets have exactly the same number of fields and with their respective
transformation laws derived from (2.1) as
$$
\eqalign{
{\d}_{Q} w_{\rm I}   &=~ i 2 \a_{\rm I} \, \xi ~-~ i 2 \, \e_{\rm{I \, J \, K}}
\a_{\rm J}  \xi_{\rm K} ~~~~, ~~~~
{\d}_{Q} \, \xi ~=~ - \, \a_I \, \pa_x w_{\rm I}   ~~~~, \cr
{\d}_{Q} \, \xi_I  &=~   \a_I \, u ~-~ \e_{\rm{I \, J \, K}}
 \a_{\rm J} \pa_x w_{\rm K} ~~~~, ~~~~
{\d}_{Q} \, u  ~=~ - i \, 2  \a_{\rm I} \, \pa_x \xi_{\rm I}
{}~~~~, } \eqno(4.8)
$$
for $N$ = 3 theory and as
$$
\eqalign{
{\d}_{Q} w_{{\rm I} \, {\rm J} }  &=~ i 2 \a_{[ {\rm I}} \, \xi_{{\rm J} ] }
 ~+~ i  2 \, \e_{\rm {I \, J \, K \, L}}  \a_{\rm K}  \xi_{\rm L} ~~~~, \cr
{\d}_{Q} \, \xi_I  &=~   \a_I \, u ~-~  \a_{\rm J} \pa_x w_{\rm {I \, J}}
{}~~~~, \cr
{\d}_{Q} \, u  &=~ - i \, 2  \a_{\rm I} \, \pa_x \xi_{\rm I}
{}~~~~. } \eqno(4.9)
$$
for $N$ = 4 theory.

Now we are exactly in the same position as with the $N$ = 2 theory.  We
begin again with (2.9).  Applying one supersymmetry variation leads to the
$N$ = 3, 4 analog of (2.10).    Applying a second supersymmetry variation
leads to the $N$ = 3, 4 analog of (2.11).  However, it is amusing to note
that our proposal for the $N$= 4 SKdV equation is almost identical in form
to the $N$ = 2 theory,
$$  \eqalign{ {~~~~~~~~~}
0 &=~  \pa_t u ~+~ \pa_x^3 u ~-~ 6 u \pa_x u  ~-~ 3 \pa_x (\,  u w_{{\rm I}
\,  {\rm J}} w_{{\rm I} \, {\rm J}} \,)  ~+~ \frac 32 \pa_x (\, w_{{\rm I}
\, {\rm J}} \pa_x^2 w_{{\rm I} \, {\rm J}} \,) {~~~~~~~~~} \cr
&{~~~}~-~ i \, 6 \pa_x (\, \xi_{\rm I} \pa_x \xi_{\rm I} \,)  ~+~  i \,
12 \,   \pa_x  (\, w_{{\rm I} \, {\rm J}} \xi_{\rm I}  \xi_{\rm J}
\,)  \ ~~~~.  }\eqno(4.10) $$

We caution the reader that the cases of $N \le $ 4 are the exception rather
than the rule. In each of these cases, we are able to formulate the
theory solely in terms of a primary submultiplet. We are able to set
the secondary submultiplet to zero as a constraint (as opposed to an
equation of motion) without disturbing the off-shell supersymmetry of
the primary submultiplet. For general values of $N$ this is not the case
and the treatment of the secondary submultiplet must handled carefully.
The simplest way to proceed is to impose the first equation in (2.9)
without taking the ``trace'' with the $f$-tensor and not use the second
equation.  Under this circumstance we are guaranteed to find a manifestly
off-shell supersymmetric system that includes the KdV equation for all values
of $N$.

The suggestion that the KdV equation admits $N = 3,4$ supersymmetric
extensions was first made in reference \cite{G1} based on the use of
harmonic superspace and superconformal algebras. It is therefore useful
to make some comparisons.  Foremost, since the off-shell structure of
our formulation in (2.1) is determined from representations of the ${
\cal G}{\cal R} ({\rm d}, N)$ algebras, we begin with a finite set of
auxiliary fields compared to the inifinite set required by harmonic
superspace. For the $N = 3$ case we seem to be in general agreement
with regard to the on-shell theory. In particular, our ${\cal U}{\cal G}
{\cal R}$-covariant formalism picks the $a = 1$ theory upon reduction
to $N = 2$.  For the $N = 4$ case we again agree with the previous
results in terms of spectrum.  However, after reduction to $N = 2$
we find only the $a = 1$ theory whereas the most recent results of
Delduc et. al. seem to suggest that the $a = 2,4$ cases are preferred.
The source of this disagreement is at present unclear.

\section{ ${\cal U}{\cal G}{\cal R}$-covariant Lax Operator}

{}~~~~As discussed previously \cite{F} by Ramos and Roca, there is a
relation between spinning particles and the Lax operator. In the following
we first review this observation briefly and make some modifications that
will be convenient later in this section.

The action for the ordinary massless relativistic particle is well known
to be described by an action that contains an einbein ($e$), momentum
($P$) and coordinate ($X$),
$$ {\cal L} ~=~ - \frac 12 \, e^{-1} P^2 ~+~ P (\, \pa_{\t} X \,) ~~~ ,
\eqno(5.1) $$
whose equations of motion follow from the calculus of variations as
$$ P^2 ~ = 0 ~~~,~~~ P ~ = ~ e \, (\, \pa_{\t} X \,) ~~~ , ~~~ \pa_{\t}
P ~=~ 0 ~~~.
\eqno(5.2) $$
Now motivated by the work of Ramos and Roca, let us perform the change
of variable described by
$$ {\Tilde X} ~\equiv ~ e^{ \frac 12} X ~~~,~~~
{\Tilde P} ~\equiv ~ e^{- \frac 12} P ~+~ \frac 12 \, e^{ \frac 12} \,
(\, \pa_{\t} \ln e \,)  X ~~~,
\eqno(5.3) $$
and concentrate on the latter two equations in (5.2). These become
$$  {\Tilde  P}  ~ = ~ (\, \pa_{\t} {\Tilde X}  \,) ~~~ , ~~~ \pa_{\t}
{\Tilde P} ~=~ - {\cal U}[e] {\Tilde X}  ~~~,
\eqno(5.4) $$
where ${\cal U}[e]$ is defined by
$$ {\cal U}[e]  ~\equiv ~ - \frac 12 \Big[ ~ (\, \pa_{\t}^2
\ln e \, ) ~+~ \frac 12 \, (\, \pa_{\t} \ln e \, )^2 ~\Big]   ~~~~.
\eqno(5.5) $$
This quantity has a number of interesting properties including
$$ {\cal U}[ {\rm J}^{-1}  ]  ~= ~ \Big( \frac {f'''}{f'} \Big) ~-~
\frac 32 \Big( \frac {f''}{f'} \Big)^2
 ~\equiv~ S(f) ~~~,
\eqno(5.6) $$
where ${\rm J} \equiv \pa_{\t} f$ is the Jacobian of the coordinate
transformation $\t \to f(\t) \equiv [\exp ( K) \t], ~K = K^{\t} \pa_{\t}$
and $S(f)$ is the Schwartzian derivative.  Also under a scale transformation
of the einbein $e \to e \exp{\l(\t)}$ (with $\l(\t)$ an arbitrary function)
we see
$$ \eqalign{
 {\cal U}[ e \exp{\l(\t)}]  &=~ {\cal U}[ e ] ~-~ \frac 12 \, ( \,
\nabla_{\t}^2 \l \, ) ~-~ \frac 14 \, ( \, \nabla_{\t} \l \,
)^2 ~~~~,  \cr
\nabla_{\t} \l &\equiv~ \pa_{\t} \l ~~~,~~~  \nabla_{\t}^2 \l ~\equiv~
( \, \pa_{\t} ~+~ (\, \pa_{\t} \ln e \, ) \, )\, \pa_{\t} \l ~~~. }
\eqno(5.7) $$
Since  ${\cal U}(1) = 0$, it follows that we can also write
${\cal U}(e) = - \frac 12 \, ( \, \nabla_{\t}^2 \ln e \, ) ~+~
\frac 14 \, ( \, \nabla_{\t} \ln e \, )^2$.

Clearly, the equations of (5.4) for ${\Tilde X}$ and ${\Tilde  P}$ are
derivable from the Hamiltonian
$$ {\cal H} ~=~ \frac 12 [~ {\Tilde  P}^2 ~+~ {\cal U}[e] {\Tilde  X}^2
  ~] ~~~~.
\eqno(5.8) $$
by use of a standard Poisson bracket. Finally the second order operator
form of (5.4) becomes
$$ \Big[ ~ \pa_{\t}^2 ~+~ {\cal U}[e] ~\Big] {\Tilde X} ~=~ 0 .
\eqno(5.9) $$
and this is the Lax operator (after we switch $\t \to x$). In \cite{F} this
argument has been extended to the case of the $N$ = 1 spinning particle
and the $N$ = 1 SKdV system.  We should be able to find a ${\cal {UGR}}
$-covariant formulation by embedding the component results into superfield
equations involving the spinning particle superfields in a manner that is
independent of N and make the switch $ \t \to x$ at the end.

We can easily embed these results into superfield equations. For example,
(5.3) can be seen to occur as components of the equations
$$
{\Tilde \Pi}_{\rm I} ~=~ E^{-\frac 12} \, \Pi_{\rm I} ~+~ i \frac 12
\, E^{\frac 12} \, \Big( D_{\rm I} \ln E \Big) X  ~~~~,~~~~ {\Tilde X}
{}~=~  E^{\frac 12} \, X  ~~~~.
\eqno(5.10) $$
Here $X$ and $\Pi_{\rm I}$ denote superfields whose component formulation
is described in reference \cite{H} as well as in the appendix.  The spinorial
derivative $D_{\rm I}$ due to (3.1) satisfies
$$
[~ D_{\rm I} \, ,\, D_{\rm J} ~\} ~=~ - i\, 4 \,  \d_{\rm {I \, J}}
\, \pa_{\t} ~~~~,
\eqno(5.11) $$
and $E$ denotes the superdeterminant of the 1D supergravity vielbein that
satisfies,
$$ [ ~ E_{\rm I} \, , \, E_{\rm J} ~ \} ~=~ - i 4 \, \d_{\rm {I
\, J}} \, E_{\t} ~~~,~~~ [ ~ E_{\rm I} \, , \,
E_{\t} ~ \} ~=~ 0 ~~~.
\eqno(5.12) $$
Similarly, the results in (5.4) can be embedded into the following
superfield equations
$$ \eqalign{
{\Tilde \Pi}_{\rm I} &=~ i \frac 12 D_{\rm I} \, {\Tilde X} ~~~~,~~~~
\pa_{\t} {\Tilde \Pi}_{\rm I} ~=~ -  i \frac 12 {\cal U}_{\rm I} [ E ] \,
{\Tilde X} ~~~~, \cr
{\cal U}_{\rm I} [ E ] ~\equiv ~ &- \frac 12 \, \Big[ ~ \Big( \pa_{\t}
D_{\rm I}   \ln E \Big) ~+~ \frac 12 \, \Big( D_{\rm I} \ln E \Big)
\, \Big( \pa_{\t} \ln E \Big) ~ \Big] ~~~~. }
\eqno(5.13) $$
Not surprisingly we find
$$ \eqalign{ {~~~~}
{\cal U}_{\rm I} [ E  \exp \L ]  &=~ {\cal U}_{\rm I} [ E ] ~-~ \frac 14
\Big[  \, ( \nabla_{\t} D_{\rm I} \L \,) \,+\, ( \nabla_{\rm I} \pa_{\t}
\L \,)  \,+\, (D_{\rm I} \L \,) \, (\pa_{\t} \L \,) \, \Big]  ~~~,  \cr
( \nabla_{\t} D_{\rm I} \L \,) \, \equiv {\,}& \Big[ \, \pa_{\t} \,+\, \Big(
\pa_{\t} \ln E \Big) \, \Big] (D_{\rm I} \L\, )  ~~, ~~ ( \nabla_{\rm I}
\pa_{\t} \L \,)  \, \equiv \,  \Big[ \, D_{\rm I} \,+\, \Big( D_{\rm I}
 \ln E \Big)  \, \Big] (\pa_{\t} \L\, )  ~~~, \cr
S(K) &\equiv~  {\cal U}_{\rm I} [ {\rm J}^{-1} ] ~~~, ~~~ {\rm J} ~\equiv~
( 1 \cdot e^{ \bvec K } ) ~~~ ,~~~ K ~\equiv~ K^{\rm I} D_{\rm I}
{}~+~ K^{\t} \pa_{\t} ~~~,
} \eqno(5.14) $$
where on the last line above we have expressed the super-Schwartzian in
terms of the super-Jacobian of the coordinate transformation induced
by the exponentiation of the super-vector field $K$ (i.e. the transformation
$(\z^{\rm I}, \, \t) \to e^K \, (\z^{\rm I}, \, \t)$).

Combining the first two equations of (5.16) we obtain
$$
\Big\{ ~ \pa_{\t}  D_{\rm I} ~+~ {\cal U}_{\rm I} [ E ] ~ \Big\}
{\Tilde X} ~=~ 0
{}~~~~,  \eqno(5.15) $$
as the ${\cal {UGR}}$-covariant generalization of (5.9).  For the case of
N = 1, the operator in this equation is precisely the super-Lax operator of
reference \cite{F}. Since this last equation is ${\cal {UGR}}$-covariant,
we propose that its interpretation as the super-Lax operator should
extend to all N.

\section{Conclusion}

{}~~~~One of the interesting points regarding supersymmetric systems
is the proposal of De Crombrugghe and Rittenberg that states that
all supersymmetric systems with $N~ >$ 4 supersymmetry must necessarily
be integrable systems. With this as a background it is not
surprising that our proposal for the $\aleph_0$ supersymmetric
extension of the KdV equation should be made.  However, we {\underline
{emphasize}} that we have {\underline {not}} given a proof that
the system of equations in (2.9) (or (6.1) below) are completely
integrable.

We believe that our results are robust.  In fact it is tempting to
conjecture that the multiplet of (2.1) is universal for {\underline {all}}
supersymmetric integrable systems in the sense that it provides the basic
supersymmetry representations for these theories.  We should mention that
there are lots of embeddings of the equations of integrable systems into
$\aleph_0$-extended systems.   What seems fairly unique about (2.1) are
the close relations to both spinning particle and 3D $\aleph_0$-extended
supersymmetric Yang-Mills theory.

It is not such a great leap to propose that other integrable systems
are amenable to a similar treatment. For example, we propose that
the Kadomtsev-Petviashvili (KP) system \cite{I} works much the same way and
for exactly the same multiplet. We begin by now assuming that each field
in the multiplet of (2.1) depends on bosonic variables $(x,y,t)$ but we
use exactly the same set of transformation laws. The only difference is
to replace the first equation in (2.9) by
$$ 0~=~
\left( f_{{\rm I} \, {\rm J}} \right)_j {}^i \Big[ ~ \pa_y^2 \, {w}_i
{}^j ~+~ \pa_x [ \,  \pa_t \, w_i {}^j ~+~ \pa_x^3 w_i {}^j
 ~+~ 6 {\rm d}^{-1}  w_k {}^l w_l {}^k \pa_x w_i {}^j ~-~ 3 \,
\pa_x ( u w_i {}^j ) \, ]~ \Big] ~~~~,
\eqno(6.1) $$
while modifying (dropping where appropriate) the second equation of (2.9).
Once again if one studies the case of $N$ = 2 utilizing the parametrization
in section 2, then (6.1) is found to contain the $N$ = 1 theory \cite{I}
as well as the proposal for the $N$ = 2 \cite{J} theory.

Our present results suggest a number of interesting departures for the
future.  Foremost, there is the issue of the rigorous proof of
integrability for the $\aleph_0$ supersymmetric models.  Should this
prove to be the case, then an interesting situation develops.
The $\aleph_0$-extended supersymmetric integrable systems are
embedded in 3D $\aleph_0$-extended supersymmetric Yang-Mills theories.
In particular 3D $\aleph_0$ supersymmetric Chern-Simons theories
(possibly coupled to matter) might then provide a universal starting point.
Such an approach would begin with $\aleph_0$ supersymmetric non-Abelian
multiplets similar to (2.2) coupled to $\aleph_0$ supersymmetric
scalar multiplets \cite{G,I1} in such a way that the spin-1 field
strength is algebraically related to currents constructed from the
matter scalar multiplets.  This constitutes an equation of motion
for a Chern-Simons matter-coupled system.   With this possibility realized,
we might be able to construct an elementary proof that the supersymmetric
version of the Atiyah conjecture is {\underline {false}}. The key point is
that 4D self-dual supersymmetric Yang-Mills theories when reduced to
3D can {\underline {never}} produce via simple mechanisms any theories
that possess more than $N$ = 8 supersymmetry!  Thus, it appears that the
role of 4D self-dual theory as originally envisioned by Atiyah might be
taken over instead by 3D Chern-Simons and supersymmetric Chern-Simons
theory as the universal generators of all integrable systems.

\noindent {\it Acknowledgment}: \newline
The authors wish to thank T. H\" ubsch and H. Nicolai for useful comments.
\newline ${~~}$\newline

\noindent
{\bf {Appendix A: Explicit ${\cal GR} (4,4)$ and ${\cal GR} (4,3)$
Representations}}

As examples of the explicit form of the ${\cal {GR}}$ matrix representations
used in the text, we present here explicit results. In the ${\cal GR} (4,3)$
case we define
$$\eqalign{
\left( {\rm L}_{\rm I} \right)_{i \hat k} &\equiv~ (\, i \s^{\Hat 2} \otimes
\s^{\Hat 1}, \, i {\rm I} \otimes \s^{\Hat 2}, \, i \s^{\Hat 2} \otimes
\s^{\Hat 3} )_{i \hat k} ~=~ - \left( {\rm R}_{\rm I} \right)_{\hat k i} ~~~,
\cr
\left( {\rm E}_{\rm I} \right)_{i \hat k} &\equiv~ (\, i \s^{\Hat 1} \otimes
\s^{\Hat 2}, \, i \s^{\Hat 2} \otimes {\rm I}, \, i \s^{\Hat 3} \otimes
\s^{\Hat 2} )_{i \hat k} ~~~~, \cr
\left( {f}_{\rm I} \right)_{i j} &\equiv~ (\, i \s^{\Hat 2} \otimes
\s^{\Hat 1}, \, i {\rm I} \otimes \s^{\Hat 2}, \, i \s^{\Hat 2} \otimes
\s^{\Hat 3} )_{i j} ~~~~, \cr
\left( {\rm F}_{\rm I} \right)_{i j} &\equiv~ (\, i \s^{\Hat 1} \otimes
\s^{\Hat 2}, \, i \s^{\Hat 2} \otimes {\rm I}, \, i \s^{\Hat 3} \otimes
\s^{\Hat 2} )_{i j} ~~~~.  }
\eqno(A.1) $$
Here the matrices ${f}_{\rm I}$ are related to the usual ${f}_{\rm I
\rm J}$-matrices via the equation
$$ {f}_{\rm I \, \rm J} ~=~ \e_{\rm I \, \rm J  \, \rm K}   {f}_{\rm K} ~~~.
\eqno(A.2) $$
We note that explicit expressions for  $\left( {\rm L}_{\rm I} {\rm E}_{
\rm J} \right)_{i \hat k} $ and $\left( {f}_{\rm I} {\rm F}_{\rm J} \right
)_{i j}$ follow from the matrix maultiplications
$$
\left( {\rm L}_{\rm I} {\rm E}_{\rm J} \right)_{i \hat k} ~=~ \left( {\rm
L}_{\rm I} \right)_{i \hat l}  \left({\rm I} \right)_{l \, \hat l}   \left(
{\rm E}_{\rm J} \right)_{l \hat k} ~~~,~~~  \left( {f}_{\rm I} {\rm F}_{\rm J}
\right)_{i j} ~=~ \left( {f}_{\rm I} \right)_{i l}  \left( {\rm F}_{\rm I}
\right)_{l \, j} ~~~~, \eqno(A.3)$$
respectively.

In the case of ${\cal GR} (4,4)$ we define,
$$\eqalign{
\left( {\rm L}_{\rm I} \right)_{i \hat k} &\equiv~ (\,
{\rm I}  \otimes  {\rm I},  \, i \s^{\Hat 2} \otimes
\s^{\Hat 1}, \, i {\rm I} \otimes \s^{\Hat 2}, \, i \s^{\Hat 2} \otimes
\s^{\Hat 3} )_{i \hat k} ~=~ - \left( {\rm R}_{\rm I} \right)_{\hat k i}
{}~~~, \cr
\left( {\rm E}_{\Hat {\rm I}} \right)_{i \hat k} &\equiv~ (\, i \s^{\Hat
1} \otimes
\s^{\Hat 2}, \, i \s^{\Hat 2} \otimes {\rm I}, \, i \s^{\Hat 3} \otimes
\s^{\Hat 2} )_{i \hat k} ~~~~.  }
\eqno(A.4) $$
The explicit forms of the matrices $\left( {f}_{\rm I \rm J} \right)_{k}
{}^l$ and $\left( {\Tilde f}_{\rm I \rm J} \right)_{\hat k} {}^{\hat l}$
follow from the definitions below equation (3.1). For the matrices denoted
by ${\rm E}_{\Hat {\rm I}}$ and ${\rm F}_{\Hat {\rm I}}$, we simply use
exactly the same matrices as for the case of N = 3.
$${~~~}$$

\noindent
{\bf {Appendix B: ${\cal GR} ({\rm d},N)$ Off-Shell Spinning Particle
Supermultiplets}}

In this appendix, we simply include the component level description of
the multiplets required to describe the off-shell spinning particle. First
there is a supermultiplet that contains the coordinate $X$. The complete
multiplet and transformation laws are given by,
$$
\eqalign{
{\d}_{Q} \, {\rm X} &=~ i  \a^{\rm I} \,   \Psi_{\rm I} ~~~, \cr
{\d}_{Q} \, \Psi_{\rm I} &=~ -2\, [~ \a_{\rm I} \, (\pa_{\t} {\rm X}) ~+~
 d^{-1} \a^{\rm J} (f_{{\rm I}\, {\rm J}})_i {}^j {\cal F}_j {}^i ~]  ~~~, \cr
{\d}_{Q} \,  {\cal F}_i {}^{\, j}  &=~ i  \a^{\rm I} \, (f_{{\rm I}\,
{\rm K}})_i {}^j  (\pa_{\t} \Psi_{\rm K} ) ~+~ i
\a^{\rm K} \, ({\rm L}_{\rm K})_i {}^{\hat k} \L_{\hat k} {}^j   ~~~ , \cr
{\d}_{Q} \, \L_{\hat k} {}^j &=~ 2 \a^{\rm K} \,\pa_{\t} \, [~ ({\rm
R}_{\rm K})_{\hat k} {}^l {\cal F}_l {}^{\, j} ~+~ d^{-1} ({\rm R}^{\rm I})_{
\hat k}{}^j (f_{{\rm I}\, {\rm K}})_k {}^l  {\cal F}_l {}^{\, k} ~] ~~~, }
\eqno(B.1)  $$
where the algebraic restrictions ${\cal F}_{i} {}^i = \left( {\rm L}_{\rm
I} \right)_j {}^{\hat k} \L_{\hat k} {}^j = 0$ are imposed.

Next there is a second supermultiplet that contains the canonically
conjugate momentum ${\rm P}$. The complete multiplet and transformation
laws take the forms,
$$ \eqalign{ {~~~~~~~~~}
{\d}_{Q} \, \p_{\rm I}  &=~   \a_{\rm I} \, {\rm P} ~+~ {\rm d}^{-1}
\a_{\rm K} \left( f_{{\rm K}\,{\rm I}}
\right)_j {}^{i} \, {\cal G}_{i} {}^{j} ~~~~, \cr
{\d}_{Q} \, \m_i {}^{\hat k} &=~ -  \, \a_{{\rm K}} \left({\rm L}_{\rm K}
\right)_{k} {}^{\hat k}  \, {\cal G}_{i} {}^{k} \, + \,  {\rm d}^{-1}
\a_{\rm K} \left( {\rm L}_{\rm I} \right)_i {}^{\hat k} \, \left( f_{{\rm I}\,
{\rm K}} \right)_k {}^{l} \, {\cal G}_{l} {}^k  ~~~~, \cr
{\d}_{Q} \, {\rm P} &=~ -i \, 2 \a_{\rm I} \, {\pa}_{\t} \p_{\rm I}
{}~~~~, \cr
{\d}_{Q} \, {\cal G}_{i} {}^{j} &=~ - i \, 2  \, [~ \a_{\rm J}
\left( f_{{\rm I}\,{\rm J}} \right)_{i} {}^{j} \, {\pa}_{\t} \p_{\rm I}
{}~+~  \a_{\rm K} \left( {\rm R}_{\rm K} \right)_{\hat k} {}^{j} \, {\pa}_{\t}
\m_i {}^{\hat k} ~]  ~~~~,}
\eqno(B.2) $$
where the algebraic restrictions ${\cal G}_{i} {}^i = \left({\rm R}_{\rm I}
\right)_{\hat k} {}^i \m_i {}^{\hat k} = 0$ are imposed.

\newpage

\end{document}